\documentclass[twoside]{article}
\usepackage[a4paper]{geometry}
\usepackage[utf8]{inputenc} % ou \usepackage[latin1]{inputenc}
\usepackage[T1]{fontenc} % ou \usepackage[OT1]{fontenc}
\usepackage[numbers]{natbib}
\usepackage{RR}
\usepackage{hyperref}
\usepackage{amsfonts}
\usepackage{amsmath}
\usepackage{amssymb}
\usepackage{amscd}
\usepackage{xspace}

\newcommand\ie{\emph{i.e.}\xspace}
\newcommand\eg{\emph{e.g.}\xspace}
\newcommand\wrt{\emph{w.r.t.}\xspace}
\RRNo{9144}
\RRdate{January 2018}
\RRauthor{% les auteurs
Pierre Blanchard\thanks[fn1]{BIOGECO, INRA, Univ. Bordeaux, 33610 Cestas, France}%
  \thanks[fn2]{Pleiade team - INRIA Bordeaux-Sud-Ouest, France}%
\and Philippe Chaumeil\thanksref{fn1}\thanksref{fn2}%
\and Jean-Marc Frigerio\thanksref{fn1}\thanksref{fn2}
\and Fr\'ed\'eric Rimet\thanks{UMR Carrtel, INRA, Thonon-les-Bains, France}
\and Franck Salin\thanksref{fn1}\thanksref{fn2}
\and Sylvie Th\'erond\thanks{IDRIS, CNRS, Orsay, france}
\and Olivier Coulaud\thanks{HiePACS team, Inria Bordeaux-Sud-Ouest, France}
\and Alain Franc\thanksref{fn1}\thanksref{fn2}\thanks{Corresponding author, \texttt{alain.franc@inria.fr}}}
\authorhead{Blanchard \& Franc \& others}
\RRtitle{Caractérisation géométrique de la biodiversité : passage à l'échelle en métagénomique}
\RRetitle{A geometric view of Biodiversity: scaling to metagenomics}
\titlehead{Geometric view on Biodiversity}
\RRresume{
Nous avons conçu un algorithme de réduction de la dimension pour explorer de nouvelles voies pour une caractérisation précise de la biodiversité, ici par une approche géométrique, qui satisfait aux critères de passage à l'échelle pour les jeux de données produits par NGS (actuellement $10^5$ reads). Cette approche est basée sur la technique dite "Multidimensional Scaling", qui permet de projeter les éléments à étudier sur un ensemble de $n$ points dans un espace euclidien de faible dimension, connaissant leurs distances respectives. Nous avons calculé toutes les distances deux à deux entre reads d'un échantillon environnemental, réalisé une MDS du tableau de distances, et analysé les projections sur les premiers axes par des techniques de visualisation. Nous avons abordé la question de la complexité quadratique du calcul des distances deux à deux en réalisant les calculs dans un Centre National disposant d'une machine hyperparallèle (Turing, une IBM Blue Gene Q), et la complexité cubique de la décomposition spectrale dans la MDS en utilisant un algorithme  de projection aléatoire dense. Nous avons appliqué cette procédure à un jeu de $10^5$ reads d'un échantillon environnemental de diatomées du lac Léman. L'analyse de la forme du nuage de points obtenu ouvre la voie vers une analyse géométrique de la biodiversité, et une construction rigoureuse d'OTUs (Operational Taxonomic Units) lorsque le jeu de données est trop grand pour mettre en oeuvre les méthodes de classification ascendante hiérarchique, non supervisée.
}

\RRabstract{
We have designed a new efficient dimensionality reduction algorithm in order to investigate new ways of accurately characterizing the biodiversity, namely from a geometric point of view, scaling with large environmental sets produced by NGS ($\sim 10^5$ sequences). The approach is based on Multidimensional Scaling (MDS) that allows for mapping items on a set of $n$ points into a low dimensional euclidean space given the set of pairwise distances. We compute all pairwise distances between reads in a given sample, run MDS on the distance matrix, and analyze the projection on first axis, by visualization tools. We have circumvented the quadratic complexity of computing pairwise distances by implementing it on a hyperparallel computer (Turing, a Blue Gene Q), and the cubic complexity of the spectral decomposition by implementing a dense random projection based algorithm. We have applied this data analysis scheme on a set of $10^5$ reads, which are amplicons of a diatom environmental sample from Lake Geneva. Analyzing the shape of the point cloud paves the way for a geometric analysis of biodiversity, and for accurately building OTUs (Operational Taxonomic Units), when the data set is too large for implementing unsupervised, hierarchical, high-dimensional  clustering.}
\RRmotcle{Biodiversité, Métabarcoding, Multidimensional Scaling, Décomposition en Valeurs Singulières, Projection aléatoire}
\RRkeyword{Biodiversity, metabarcoding, Multidimensional Scaling, Singular Value Decomposition, Random Projection}
\RRprojets{Pleiade \& HiePACS}
\RCBordeaux % Sophia Antipolis M\'editerran\'ee

\begin{document}
\makeRR   % cas d'un rapport de recherche

\newpage
\tableofcontents

\vspace*{2cm}

Deciphering the diversity of life has been a longstanding guiding thread in biology \cite{Mayr1982}. Diversity relies on differences and dissimilarities. A community is the set of organisms sharing a common habitat \cite{Ricklefs2000}. There has been numerous definitions of the diversity of a community, excellently reviewed in \cite{Margurran2003}. Biodiversity is not random, but organized in a given number of patterns \cite{Heywood1995}. Identifying those patterns shaped by evolutionary forces is one of the key challenges in  biology \cite{Levin1992,Gaston2000}. All these approaches rely on the simplification of the whole diversity as an index, be it the Shannon index, the Simpson index, or even the number of different species. A first objective of this work is to characterize diversity as the shape of a point cloud, where a point is an organism in a community, and localized at a distance from neighbors which represents their dissimilarities. Such an approach is called here the \emph{geometric view on biodiversity} because a unique index or a set of indices is replaced by the shape of a point cloud. Rigorously defining the shape of a point cloud is not an easy task, intertwining computer vision and machine learning \cite{Szeliski2011}. Here, we address a simplification of this question, focusing on one type of shapes which are classical in numerical taxonomy \cite{Sneath1973}: organized sets of clusters. A second objective of this paper is to propose a connection between High Performance Computing and molecular based numerical taxonomy, in order to handle with highest accuracy all the information provided by very large datasets produced by NGS facilities. This connection is established by both using massive parallelization and providing efficient algorithms for linear algebra behind dimensionality reduction of massive data.

\section{Introduction}
\label{sec:intro}

%\subsection{Molecular taxonomy}
\paragraph{Molecular based taxonomy:} For a couple of centuries, the diversity of life has been organized as taxonomic systems, based on nested clustering \cite{Sneath1973}. This relied mainly as an expertise on key dissimilarities between 
some phenotypic traits, and systematics was a cornerstone of natural history. More recently, systematics has shifted towards molecular systematics \cite{Hillis96}, where 
distances between sequences have been organized along phylogenies, which model the history of speciation in a given group \cite{Felsenstein2004}. Slightly later, so called 
barcoding has been  an international effort to normalize sequencing (primers choice) to encompass as much as possible of the diversity of all organisms, using very large and 
shared databases \cite{hebert03,Hollingsworth2009,Pawlowski2012}. Such an approach is perfectly suited for the geometric view of biodiversity as a replicable procedure, because 
there exists algorithms for computing distances between two sequences. In this work we focus on a molecular based construction which scales with massive data sets produced by NGS. Diversity characterization is obtained by visualization of the produced point clouds by projection in low dimension spaces, 
as finding a shape in a high dimensional space still remains an open question.
%\subsection{Metabarcoding}
\paragraph{Metabarcoding:} As soon as molecular diversity has been put in relation with taxonomic diversity, several studies have shown that most of the existing diversity had been hidden to our eyes. This 
is particularly true for microbial diversity in the oceans \cite{Lopez-Garcia2001,Sogin2006}. In parallel, metagenomics has emerged as a technique to sequence simultaneously 
all genomes in a given microbial community. Such a technique has been developed as metabarcoding  \cite{Hajibabaei2011,Bik2012,Taberlet2012,Kermarrec2014,Dafforn2014,Joly2014,Pawlowski2014} by amplifying and sequencing all the sequences of a given marker of taxonomic interest in a given community. Here, we develop some data analysis on metabarcoding of environmental samples of freshwater diatoms. We have selected this clade as it is one of the clades of microbial organisms for which an expertise exists for optical based taxonomic identification \cite{Mann1999}. Hence, consistency between morphological based and molecular based taxonomy can be evaluated.

\section{Data analysis}\label{sec:datanalysis}

As in machine learning, two main families of methods exist to assign taxonomically a set of sequences:
\begin{itemize}
 \item supervised learning, where an unknown sequence is assigned to a taxon by comparison with a reference database, where sequences have been taxonomically annotated
 \item unsupervised learning, where the organization of the diversity itself is sought for, as in clustering. This is traditionally called OTU picking in metabarcoding studies \cite{Bik2012}.
\end{itemize}
There exists classical and efficient tools for each of this task. BLAST \cite{Altschul1990} is by far the most used tool for supervised learning. However, there is no equivalent golden standard for unsupervised clustering. It should be nested agglomerating clustering, which suffers from computation load as computing all distances between $n$ sequences and performing agglomerative clustering are each quadratic with $n$, i.e. the complexity is in $\mathcal{O}(n^2)$ \cite{Muellner2013}. Hence, many heuristics have been used. Among them, one of the mostly used is an adaptation of k-means \cite{Edgar2010}. It can handle huge datasets, as it can work in streaming, but it suffers from some flaws, as the fact that the result depends on the order the sequences have been presented. Swarm \cite{Mahe2014} has solved some of these issues, but still relies on some heuristics. However, there is no better tool for huge data sets (millions of sequences) for clustering. Here, we compute exact distances between sequences, and run Multidimensional Scaling as a way to visualize the structure of the diversity as given by pairwise distances.\\
\\
Historically, Multidimensional Scaling has been derived to find configurations of points in Euclidean space of small dimension the shape of which fits as much as possible a set of pairwise distances \cite{Torgerson1952}. However, the machine learning algorithms used for building a point cloud knowing pairwise distances, as well as just computing the distances between all pairs of reads, have a 
complexity respectively cubic and quadratic to the number of specimens. Such a complexity has been acknowledged as a difficulty for dealing with full matrices of pairwise distances, and several procedures have been designed to circumvent it (see \cite{Aflalo2013} and references therein). In this work, we implement an algorithm to circumvent the complexity of the problem by computing pairwise distances on an hyperparallel supercomputer, and by calculating the desired eigenpairs by methods based on random projection \cite{Vempala2004,Halko2011}.

%\newpage
\subsection{A generic approach for large biological datasets}
\label{sec:generic-approach}
% non-metric dissimilarity

	%\subsection{Pipeline}
	%\label{sub:pipe}
	%
	The pipeline we have designed and implemented is the sequence of following steps:
	\begin{itemize}
	 \item compute pairwise distances between all pairs of $n$ reads,
	 \item run Multidimensional Scaling on the distance matrix,
	 \item visualize the first $r$ components with $r \ll n$.
	\end{itemize}
	The novelty lies in an implementation on large data sets, \ie, scaling up to data sets produced by NGS sequencing operations, currently $n \simeq 10^5$ reads. Computing pairwise distances between reads is quadratic in $n$, and requires $\mathcal{O}(Kn^2)$ elementary operations, $K$ being a constant quadratic in the length of the reads. Running MDS requires computing $r$ eigenpairs of a $n \times n$ matrix, which can be done in $\mathcal{O}(rn^2)$ operations. We address this complexity issue by 
	\begin{itemize}
	 \item parallelization of the computation of pairwise distances on an hyperparallel machine (an IBM Blue Gene Q)
	 \item using a MDS approach powered by random projection and efficient C++ implementation.
	\end{itemize}

	% Read distance matrix 16*0.30s (16*1s at first execution)
	% Write 10k random reads 0.6s
	\noindent The time needed for each of these steps is given in table \ref{tab:mds-chain}.
	\begin{table}[htbp]
	  \centering
	  \footnotesize
	  \begin{tabular}{c | c | c | c | c}
	      Sample      & compute distances    & transfer and convert                    & subsample    & perform MDS  \\ %[0.5ex] %
	    (Lake Geneva) & S-W on \emph{Turing} & \emph{iRods}$\rightarrow$\emph{plafrim} & $10^4$ reads & full SVD     \\ %[0.5ex] %
        \hline
	    $10^5$ reads  &   4h                 &     20min $+$ 1h40                      &    30s         &   20min     \\ %[1ex] %adds vertical space
	  \end{tabular}
	  \caption[Estimation of the time required to perform expensive computations in our operational workflow.]{Some average running times related to expensive operations involved in the operational chain. %\underline{Observations:} The large amount of time required to convert distance matrices from text files to binaries actually represent a significant gain compared to the initial state of the chain (text files read from python). In fact, since operations such as subsampling may be repeated numerous times for a given sample, it is recommended to store the full distance matrix on the disk in a format that allows for a fast loading of the full sample in memory.
	  }
	  \label{tab:mds-chain}
	\end{table}

%\newpage
\subsection{Scalable pairwise distances computation}
\label{sub:dis}
	% S-W algorithm
	% Property of dissimilarity: distance? resolution?
	There are several ways to compare sequences (see e.~g. \cite{gusfield97,Yang2006}). The dissimilarity used here is computed from a local alignment score \cite[p.~232]{gusfield97} 
	using Smith-Waterman algorithm \cite{SW81}. If the length of the sequences is $p$, the algorithm scales like $\mathcal{O}(p^2)$. As it is quadratic in the number $n$ of sequences, computing the 
	whole matrix has a complexity of $\mathcal{O}(n^2p^2)$. We have written a program in C, called \texttt{disseq}, which takes as inputs two fasta files, of length $m$ and $n$ each, and returns the $m \times n$ matrix $D=[d_{ij}]$ where $d_{ij}$ is the distance between sequence $i$ of first file and $j$ of second file. This program has been parallelized with MPI 
	(Message Passing Interface) as a map-reduce process, in a program called \texttt{mpi-disseq}, as all distances can be computed independently. \texttt{mpi-disseq} has been run 
	on a BlueGene Q (IBM) located at IDRIS to compute the matrix $D$ of all pairwise distances ($\sim 5.10^{9}$ values). Its architecture is made of 6 racks, of 1,024 nodes each, with 16 cores per node, hence 98,304 cores. We have used Turing with $2^{14} = 
	16,384$ cores (one rack). Turing peak power is 1,258 Pflops/s. One advantage of such a choice, beside massive parallelization, is a low energy consumption. Such an architecture 
	is particularly suitable for massive embarrassingly parallel jobs. The program has been tested on Avakas (Mésocentre de Calcul Intensif Aquitain, 264 computing nodes, 12 cores per 
	node), and then ported to Turing. It scales perfectly.

%\newpage	
\section{A random projection-based spectral decomposition}
\label{sec:randproj}
In this section, $D$ denotes the pairwise distance matrix and $G$ the Gram matrix built from it (see section \ref{sec:mm}). Next step is to perform MDS on $D$. The calculation is presented in section Material and Methods. It amounts to perform a SVD of a Gram matrix $G$ built from the distance matrix $D$, and of same dimension. In this section we describe an efficient approach for computing the spectral decomposition of the real symmetric matrix $G$ at a quadratic cost in $n$. We put the emphasize on the genericity of the approach and its competitive numerical performance (see \cite{Blanchard2017} for details). 
\subsection{Standard iterative approach}
\label{sub:randproj-std}

The standard methods used to compute the full eigenvalue decomposition (EVD) of an arbitrary $n$-by-$n$ real matrix usually take $\mathcal{O}(n^3)$ operations. This complexity makes those methods computationally untractable to large data sets. However, there exists well-known iterative techniques that compute the first $r$ eigenpairs at a $\mathcal{O}(n^2r)$ cost, \eg, Arnoldi or Lanczos algorithm \cite{Sorensen1997}. Despite presenting several flaws, these variants provide an exact spectral decomposition.
%In order to make the MDS algorithm work, the matrix needs to be symmetric positive semi-definite. %Therefore, eigenvalues and singular values are related by $\lambda = \sigma^2$, and we may as well use a random projection algorithm to compute singular values and singular vectors to build $X$. 

\subsection{A \emph{random projection}-based algorithm}
\label{sub:randproj-principle}
% A randomized algorithm
% Show no algo
Our novel contribution to MDS consists in using low-rank approximation techniques based on random projection, to compute an approximate rank-$r$ spectral decomposition of $G$. 
Random projection, along with random sampling, belongs to a broader class of dimensionality reduction tools known as random sketching \cite{woodruff2014sketching}. Although most fast approaches to MDS rely on random sampling \cite{platt2005fastmap}, to our knowledge there are very few contributions to this field that involve random projection if any. Despite involving more intensive computations than random sampling, random projection presents many benefits such as better accuracy, robustness and low variability. %Finally, since the efficiency of the randomized SVD mostly depend on the ability to apply fast matrix multiplication, it can be improved by exploiting the structure of the matrix and it can be parallelized straightforwardly.
Low-rank approximation techniques based on random projection were made popular by a SIAM review paper by Halko et al. \cite{Halko2011}. Among theses algorithms, the randomized Singular Value Decomposition (or randomized SVD) computes an approximate rank-$r$ SVD of a $m$-by-$n$ matrix in $\mathcal{O}(mnr)$ operations. 
This algorithm can be used to compute the first eigenpairs of $G$, by simply remembering that for a symmetric real matrix singular values $\sigma$ can be related to the eigenvalues by $\sigma=|\lambda|$. 

\subsection{Numerical performance}
\label{sub:randproj-numperf}
% Computational cost

This approach represents a significant alternative to standard EVD algorithms, as it shares the same complexity, namely $\mathcal{O}(rn^2)$, but usually performs significantly better and parallelizes straightforwardly. However, the input matrix has to fulfill a few criteria for the method to work properly. $G$ should be low-rank, \ie, $r \ll n$, and have a fast decreasing spectrum or a low stable-rank, \ie, $st(G)=\|G\|^2_F/\|G\|^2_S \ll n$, \ie a low ratio between the squared Frobenius norm $\|G\|^2_F=\sum_{i\leq n}\sigma_i^2$ and the squared spectral norm $\|G\|^2_S=\max_{i\leq n}\sigma_i^2$. Finally, as stated in \cite{Halko2011}, even though the algorithm has a non-zero chance to fail, tight Frobenius and spectral error bounds were shown to hold with high probability, if $G$ verifies the aforementioned conditions. %Additionally, those bounds can be significantly improved by considering oversampling or power iterations.
-
	
\section{Visualization of a large diatoms sample}
\label{sec:exp}

The main goal of our approach is to achieve an efficient and handy visualization technique that provides relevant information on the diversity of some real-life samples, currently of about $10^5$ sequences. Due to the longstanding coevolution between clustering and numerical taxonomy, a possible analysis of the shape of the point cloud consists in identifying clusters. Clustering is an immense domain, and we will not enter here into the discussion for the best methods knowing a pairwise distance matrix: this is ongoing work. We focus in this report on studying the concentration of the reads in some subsets in the projection on the first few dimensions of MDS.

\subsection{Dimensionality reduction}

When the dimension of a space is large, the so called \emph{curse of dimensionality} enters into the game. In brief, it can be summarized by telling that there is much room in a high-dimensional space. For example, two opposite vertices of an hypercube are at distance $\sqrt{n}$ (i.e. the distance between two points in a symmetric body of unit volume can be arbitrarily large), whereas the volume of the hypersphere of radius one shrinks to zero: the probability to have a neighbor at distance one or less is negligible \cite{Izenman2008,Wang2012}. This can impact any algorithm based on distances. Therefore, a first step is often to reduce the dimension of the problem, by building a mapping of the original point cloud on a Euclidean space of far lower dimension. This is called dimensionality reduction, and is a key step in machine learning \cite{Izenman2008,Lee2007,Murphy2012}. MDS is the tool for visualizing the geometry of the point cloud associated to distances between reads in low dimension. Hence we present such a visualization by projection on first axis of MDS, and provide some hints for interpretation by a comparison with supervised clustering. We do not report here on unsupervised clustering, which is deferred to further work. We have worked with a full sample named $L_6$ ($n=99,594$, see sample description in Materials and Methods section). We have built the associated point cloud by mean of a rank $r=50$ randomized SVD. Here, we show representations of the full sample in 3D for easy visualization of the point cloud. In such a representation, each point represents a read. We have used a supervised clustering technique (a program called \texttt{diagno-syst}, see \cite{Frigerio2016}) to assess a taxonomical name to each read for which it has been possible, to test whether the clusters were related to species. Then, we discuss the visualization of the cloud in higher dimensions.

\subsection{Reads concentration in low dimension}
\label{sub:exp-visu-concentration}

\begin{figure}[h]
\centering
\includegraphics[trim=.5cm .25cm .5cm .5cm,clip,width=.5\textwidth]{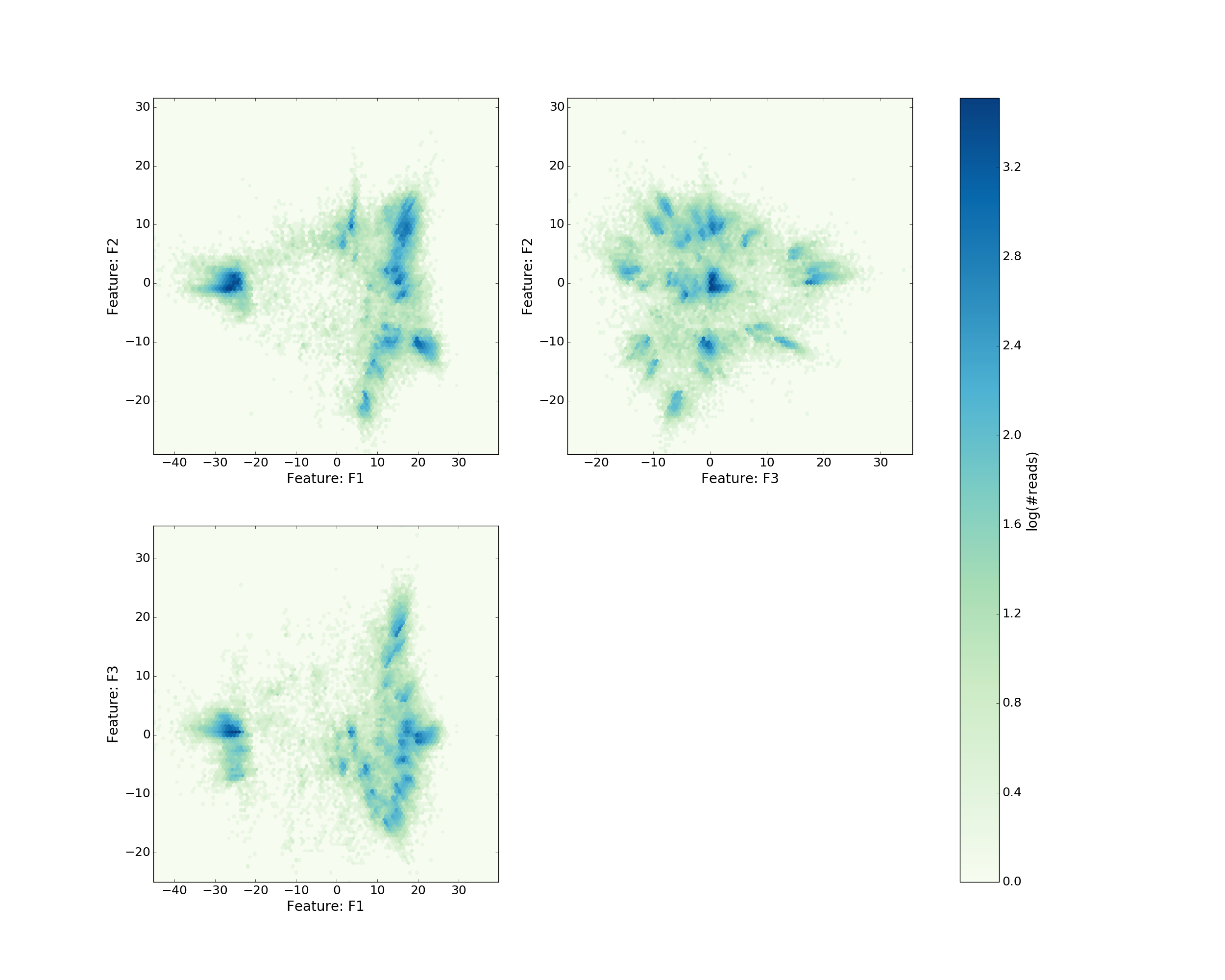}
\caption{Concentration of the population of a full sample represented in log-scale on the first thee axis.}\label{fig:L6_Full_HexBin}
\end{figure}

Because of the number of points, several reads are projected on the same pixel in regions with high concentration of reads in first axis. We have quantified this by hexagonal binning: the plane formed by axis $i$ and $j$ (with $(i,j) \in (1,2); (1,3); (2,3)$) is tessalated by hexagons. Then, the number of points per hexagon is counted, and these figures are displayed, in logarithmic scale. This is presented in figure \ref{fig:L6_Full_HexBin}. In particular, this figure shows that only very specific regions of the full domain contain most of the individuals and that they are surrounded by regions of far lower density. The pattern is an archipelago of islands of high density surrounded by an ocean of low density. It may be tempting to associate high density islands to clusters and clusters to species.
% Hex-Bin density

\subsection{Interpretation with supervised classification}
%\label{sub:exp-visu-3D}

\begin{figure}[h]
\centering
\includegraphics[trim=.5cm .25cm .5cm .5cm,clip,width=.5\textwidth]{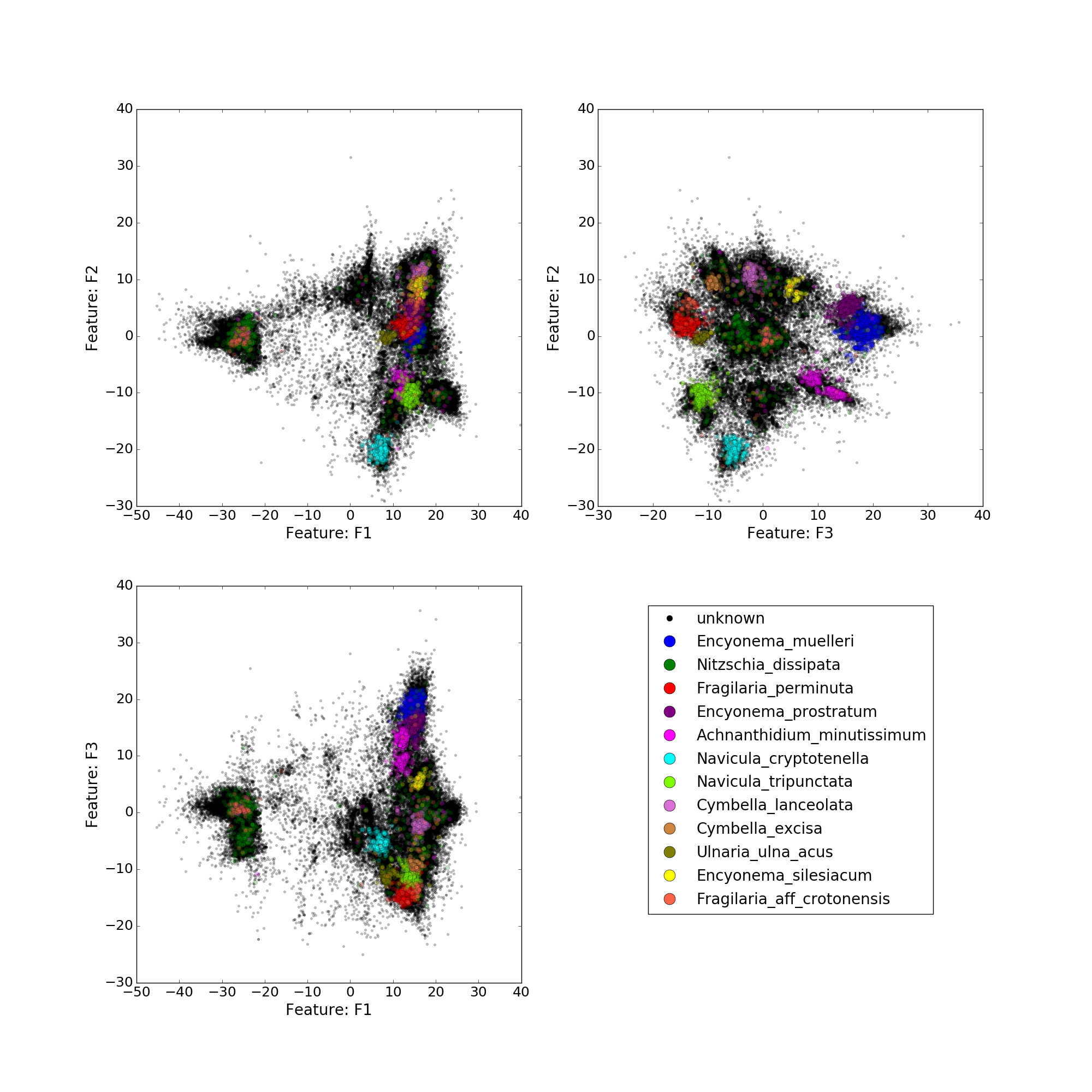}
\caption{Concentration of the population of a full sample represented in log-scale on the 3 first pairs of features.}\label{fig:L6_Full_Plot2D}
\end{figure}

In order to validate our approach, and test whether high density islands correspond to reads belonging to a same species, with ideally a one to one correspondence between islands and species, we have colored the points according to an information provided by supervised clustering on all the reads. Indeed, we have at disposal a reference database for diatoms, which is fairly good for Geneva lake \cite{Rimet2016}. We have computed all pairwise distances between the reads in our sample $(\simeq 10^5)$, and the taxonomically annotated reads in the reference database $(\simeq 2.10^3)$. This is intensive computing too, which has been executed on Turing, with \texttt{MPI-disseq} as well. Then, we have selected an homology gap (classically 97\%), and, for each read in our sample, selected all reads in the reference database which were at distance lower than the homology gap. If they all belong to a same species, the read has been annotated with this species name. If not, it has been annotated as ambiguous status. This has been done with a program called \texttt{diagno-syst}, as a work available in \cite{Frigerio2016}. Then, we have colored the individuals \wrt their species as identified by supervised clustering in different colors for different species.  (see figure \ref{fig:L6_Full_Plot2D}). We have checked as well that the species identified this way in the community have been identified optically too. This supports a consistency between morphological based and molecular based taxonomy.

% Plot2D with colors (todo opacity)

\subsection{Representation in many dimensions}
\label{sub:exp-visu-manyD}
% Represebtation using parallel coords
% Fraction of the energy of the full matrix represented on 10 components
% Discuss clustering and id species

\begin{figure}[h]
\centering
\includegraphics[width=.5\textwidth]{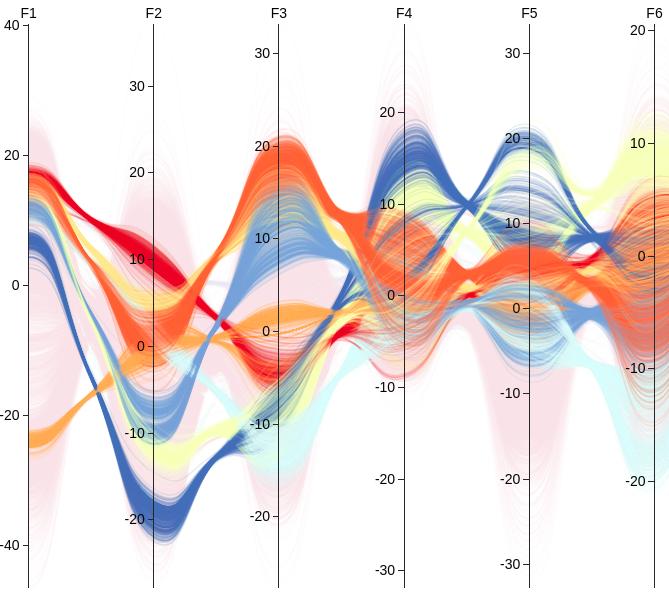}
\caption{Representation of the predominant identified species within the full sample using parallel coordinates over the first 6 features produced by random projection-aided MDS. \underline{Legend:} Encyonema muelleri (orange), Fragilaria perminuta (light blue), Nitzschia dissipita (dark yellow), Encyonema prostratum (yellow), Achnanthidium minutissimum (blue), Navicula cryptotenella (dark blue), Navicula tripunctata (light yellow), Cymbella lanceolata (red). The remaining reads appear in light red in the background.}\label{fig:L6_Full_ParaCoord}
\end{figure}

%\paragraph{Parallel Coordinates} 
In order to visualize the point cloud in more dimensions than 3D, a common alternative to 2D plots is the representation in \emph{parallel coordinates} such as the one displayed on Figure~\ref{fig:L6_Full_ParaCoord}. This technique offers various advantages as it decouples the dimensions and displays them on a single axis. Due to significant density of the cloud, we make use of advanced rendering techniques (opacity, brushing and bundling) provided by the javascript libraries \texttt{d3.js} and \texttt{paracoords.js} in order to better represent the concentration of reads. The clustering of the reads can again be confirmed by observing that individuals of the same species follow a similar path along the first few dimensions, while spreading only on a fraction of the full domain. This observation is made clearer by the observation of each species separately. The consistency of clustering as seen through bundles of trajectories of all reads belonging to the same species with species delineation is shown here up to the first ten dimensions, which permits to envisage such an approach for pattern discovery related to OTU beyond visual inspection in the first three axis. 

% Parallel-Coordinates

\section{Conclusions and perspectives}

We have implemented Multidimensional Scaling of the matrix which contained exact distances between all pairs or reads of an environmental sample of $\simeq 10^5$ reads. This has permitted to visualize the points cloud associated to those distances in low dimension spaces. We have shown that points were clustered in islands of high density regions forming an archipelago within an ocean of low density regions.  We have shown by supervised classification that some clusters could be associated to those species which were optically identified in the sample and in the reference database. This establishes consistency between supervised and unsupervised approaches, and paves the way for using unsupervised methods when taxonomic information for supervised approaches is lacking, which will greatly expand the scope of these methods.\\
\\
It can be observed as well that some islands could not be assigned to species by supervised clustering. It may be tempting to hypothesize that those clusters correspond to species which are present in the sample, but not in the reference database. More precisely, some parts of the cloud contains unknown individuals but still exhibits a certain structure and a significant density, which suggests that a deeper analysis of the cloud is possible. Supervised method permit a dictionary between molecular based patterns as derived in metabarcoding and previous patterns in biodiversity studies as derived in Natural History. However, one of the most critical shortcoming of supervised methods is that reference databases will never be complete, because of the tremendous number of species, most of them still unknown. Hence, there is an urgent need for unsupervised methods, which are consistent with supervised methods when taxonomic information is available. We have proposed here such a method by an analysis of high density regions in projection in a low dimension space. Nest step may be the use of image analysis techniques to identify and characterize those high-density islands.\\
\\
Classically, unsupervised clustering of reads for metabarcoding is done by greedy methods, mentioned in the introduction of section \ref{sec:datanalysis} (a couple of them are included as choices in Mothur \cite{Schloss2009} or Qiime \cite{Caporaso2010}). Once again, those methods rely on heuristics, and their accuracy are not assessed for very large datasets. Many different exact methods can be implemented provided one accepts to invest in computation load for a better accuracy. One can mention a few: nested aggregative clustering \cite{Muellner2013}, selecting a threshold $\theta$ and build a graph $G=(V,E)$ with $V$ being the set of reads, and $(i,j) \in E \: \Leftrightarrow \: d(i,j) \leq \theta$, and build the connected components of $G$ (this is linear in time with $n$), or more elaborate methods on the same graph like spectral clustering to derive communities \cite{UvLux07,Girvan2002}.\\
\\
Next steps are not only to compare, but to associate all these methods and tools, in order to build Operational taxonomic Unit with best accuracy on very large data sets, as consistent as possible with our knowledge of the diversity of life as we have inherited it from centuries of studies in Natural History, and as automatized and sound as possible for exploring the unknown diversity.

\section{Material and methods}\label{sec:mm}

\subsection{A sample from Lake Geneva}
\label{su-b:exp-spl}
% Describe sample
We have considered 10 environmental samples, denoted $L_t$, that were collected from Lake Geneva at about monthly intervals at times $t=1,\ldots,10$ between April 2012 and March 2013 
in order to investigate a seasonal dynamics. Amplicons of chloroplastic marker rbcL have been produced for each sample by DNA extraction, amplification, and sequencing on a Ion Torrent PGM (see \cite{Kermarrec2013} for protocols). The various samples contain from about $7\times 10^4$ to 
$1.4 \times 10^5$ reads. The diversity of each sample has been assessed both optically, and by supervised clustering of reads by mapping on a dedicated reference database (see \cite{Rimet2016} for the reference database, and \cite{Frigerio2016} for the algorithm for mapping reads, and its implementation.)
A fraction only of diversity can be assessed with these classical tools.

\subsection{Multidimensional Scaling}
\label{sub:mds}
%
%\subsubsection*{Algorithm} % Algorithm
The method implemented here is \emph{classical multidimensional scaling}\footnote{There is another procedure bearing the same name, called nonmetric MDS due to Kruskal \cite{Kruskal1964}, based on an optimization scheme, which currently cannot be implemented for more than a few thousands items.} (see \cite{Borg2005,Izenman2008} for a recent survey, and \cite{Cox2001} for a seminal monograph) which follows Torgerson's  work \cite{Torgerson1952}. Let us have a set of $n$ items with $i \in V = \{1,n\}$, and a distance $d_{ij}$ between items $i$ and $j$. Then, $(V,d)$ is a finite metric space. For a given dimension $r \in \mathbb{N}$, classical MDS is finding a map $\begin{CD} x \: : \: i @>>> x_i \in \mathbb{R}^r \end{CD}$ such that $\|x_i-x_j\|$ is as close as possible to $d_{ij}$. If the distances $d_{ij}$ are such that the corresponding Gram matrix $G$ is definite positive, the solution is well known \cite{Cox2001}, and is implemented in three steps
\begin{enumerate}
\item The scalar product $\langle x_i,x_j\rangle$ can be computed from the distances only, as
\begin{equation*}
  \langle x_i,x_j\rangle = -\frac{1}{2}\left(d_{ij}^2 - \frac{1}{n}\sum_id_{ij}^2 - \frac{1}{n}\sum_jd_{ij}^2 
  + \frac{1}{n^2}\sum_{i,j}d_{ij}^2\right)
\end{equation*}
The scalar products $\langle x_i,x_j\rangle$ are the elements of the Gram matrix $G = [\gamma_{ij}]$ with $\gamma_{ij} = \langle x_i,x_j\rangle$
\item compute the eigenvalues and eigenvectors of Gram matrix $G$ 
\begin{equation*}
Gu_\alpha = \lambda_\alpha u_\alpha, \qquad \lambda_1 \geq \ldots \geq \lambda_n \geq 0
\end{equation*}
(As $G$ is a Gram matrix, all its eigenvalues are non negative, as there is a $n \times n$ matrix $X$ such that $G=XX'$)
\item Let us denote $\Sigma = \Lambda^{1/2}$ where $\Lambda$ is the diagonal matrix with $(\lambda_\alpha)_\alpha$ on its diagonal. Then compute
\begin{equation*}
X = U\Sigma 
\end{equation*}
\end{enumerate}
Then, the best representaton of $D$ as distances within a point cloud in $\mathbb{R}^k$ embedded with standard inner product is $X_r$, which is the extraction of the first $r$ columns of $X$. $x_i$ is the $i-$th row of $X_r$. The best rank $k$ approximation of $G$ is $G_r=X_rX_r'$.

\subsection{Observations}
\label{sub:mds-obs}

%\subsubsection*{Observations} % Observations
Two observations can be made:
\begin{itemize}
\item In general, not all eigenvalues of $G$ are non negative. If all eigenvalues are non negative, then $(V,d)$ can be embedded isometrically in a Euclidean space, and a best low dimensional approximation can be computed as the Principal Component Analysis of the point cloud \cite{Mardia1979}. The condition can be seen directly on the distance matrix, and involves signs of Cayley-Menger determinents (see \cite[thrm 2.1, p. 15]{LLMM14}). 
\item  As far as calculation is concerned, step 2 is the most demanding: computing eigenpairs of the Gram matrix. We have implemented here a procedure relying on random projection (see \cite{Vempala2004} for a survey). %Such an approach is presented in subsection~\ref{sec:randproj}.
\end{itemize}

\paragraph{Acknowledgments:} Computation at IDRIS has been supported by DARI project 
\emph{Biodiversiton} \texttt{i2015037360}. Experiments presented in this paper were carried out 
using the PlaFRIM experimental testbed, being developed under the Inria PlaFRIM development action with support from LABRI and IMB and other entities: 
Conseil Régional d’Aquitaine, FeDER, Université de Bordeaux and CNRS (see https://plafrim.bordeaux.inria.fr/). Sequencing has been performed at the Genome Transcriptome Facility of Bordeaux (grants from the Conseil Régional d’Aquitaine n°20030304002FA and 20040305003FA, from the European Union FEDER n°2003227 and from Investissements d'Avenir ANR-10-EQPX-16-01) and with support of ONEMA project on Mayotte. Humid lab (DNA extraction, PCR) have been done at UMR Carrtel, under guidance of Agnès Bouchez, with support of ONEMA project Mayotte. We acknowledge an \emph{Investissement d’Avenir} grant of the Agence Nationale de la Recherche (CEBA: ANR-10-LABX-25-01). We acknowledge the nice atmosphere of network R-Syst, and support by INRA divisions EFPA and SPE, for methodological developments on metabarcoding, as well as its database infrastructure. 

\newpage

% Bibliography
\bibliographystyle{alpha}
%\bibliography{mds,references,biodiversity}
\newcommand{\etalchar}[1]{$^{#1}$}

\end{document}